\documentclass[prl,twocolumn, final, showpacs, floatfix]{revtex4}

\usepackage{amssymb,amsmath}
\usepackage{epsfig}
\usepackage{psfrag}
\usepackage{epstopdf}

\input xy
\xyoption{all}

\begin{document}

\title{On the equivalence between stochastic baker's maps
and two-dimensional spin systems}

\author{Kristian Lindgren}
\affiliation{Complex Systems Group, Department of Energy and Environment, Chalmers University of Technology, 412 96 G\"{o}teborg, Sweden}

\begin{abstract}

We show that there is a class of stochastic baker's transformations that is equivalent to the class of equilibrium solutions of two-dimensional spin systems with finite interaction. The construction is such that the equilibrium distribution of the spin lattice is identical to the invariant measure in the corresponding baker's transformation. We also find that the entropy of the spin system is up to a constant equal to the rate of entropy production in the corresponding stochastic baker's transformation. We illustrate the equivalence by deriving two stochastic baker's maps representing the Ising model at a temperature above and below the critical temperature, respectively. We calculate the invariant measure of the stochastic baker's transformation numerically. The equivalence is demonstrated by finding that the free energy in the baker system is in agreement with analytic results of the two-dimensional Ising model.

\end{abstract}

\maketitle

%\noindent{\small Keywords: baker's map, spin system, Ising model.}\\
The Ising model and the baker's map \cite{Arnold68} are two examples of canonical model systems that have provided deep insights into statistical mechanics and dynamical systems theory, respectively. The aim of this paper is to point out and illustrate a relatively simple connection between spin systems in two-dimensions and a class of stochastic baker's transformations. In this way we demonstrate a new connection between equilibrium lattice systems and low-dimensional dynamical systems.

Formal connections between dynamical systems and two-dimensional spin systems have been presented before. Examples include cellular automata (e.g., \cite{Domany84}) and coupled map lattices (e.g., \cite{Just98, Sakaguchi99}). We introduce a new class of stochastic baker's transformations and show that the noise can be constructed so that the representation of the dynamics is identical to a representation of equilibrium spin systems in two dimensions \cite{Kramers41}. The novelty in the present approach is the design of a specific family of transformations in which the invariant measures are identical to the equilibrium distributions of two-dimensional spin systems. This equivalence could potentially provide new perspectives by enabling the use of methods and results from one area in the other. A similar idea of baker's map construction has been used for establishing a connection between low-dimensional dynamical systems and universal computation \cite{Moore90}.

Consider the following stochastic baker's map
\begin{equation}
f(x,y) = (2x-\lfloor{}2x\rfloor{}, y/2 + {\xi} (x,y)/2)
\label{eq: baker map}
\end{equation}
where $\xi(x,y)$ is a stochastic variable that can take the values 0 or 1, and where $\lfloor{}x\rfloor{}$ denotes the integer part of $x$. If $\xi = \lfloor{}2x\rfloor{}$, we get the standard non-stochastic baker's map. In this paper we will consider a general case where $\xi$ is characterised by a probability distribution that depends on the position $(x, y)$ in state space. 
 
The dynamics of the baker's map is conveniently described using the dyadic representation (i.e., binary expansion) for positions in the unit interval, using the notation $x=(.x_0 x_1 x_2 ...)$ and $y=(.y_0 y_1 y_2 ...)$, so that $x = \sum_{i=0}^\infty x_i/2^{i+1}$, with $x_i \in \{0,1\}$, and similarly for $y$. The original non-stochastic map shifts the $x$ sequence to the left removing $x_0$ and the $y$ sequence to the right putting the symbol $x_0$ in the 0'th position, i.e., $x'=(.x_1 x_2 x_3 ...)$ and $y'=(.x_0 y_0 y_1 ...)$. The two semi-infinite sequences can be put together $(... y_2 y_1 y_0 . x_0 x_1 x_2 ...)$. Using this representation the map is simply an operation shifting the combined sequence one step to the left.

When we generalize the symbolic dynamics description to the stochastic baker's map, Eq.~(\ref{eq: baker map}), an iteration again means that the $x_0$ symbol is removed but that the value of the stochastic variable $\xi$ enters in the 0'th position of $y$,
\begin{equation}
f(... y_2 y_1 y_0 . x_0 x_1 x_2 ...) = (... y_1 y_0 \xi . x_1 x_2 x_3 ...)
\label{eq: bi-infinite sequence}
\end{equation}
In general we will consider the case when the probability distribution for $\xi$ depends on position $(x,y)$, i.e., the full $x$ and $y$ sequences, and we denote the probability for $\xi$ being 0 by $q_0(x,y)$.

In this paper, we are primarily interested in the case when the stationary measure of the map in the unit square is symmetric under exchange of $x$ and $y$, since such a symmetry will be required when we make the connection to spin systems below. One way to achieve this is to consider a map $\phi (x,y)$ that randomly chooses, with equal probabilities, either the function $f(x,y)$ in Eq.~(\ref{eq: baker map}) or the function $f(y,x)$.
\begin{equation}
\phi (x,y)=\left\{ {f(x,y)\ \text{with probability 1/2} \atop f(y,x)\ \text{with probability 1/2}}\right.
\label{eq: mirror map}
\end{equation}
The stochastic baker's map $\phi$ is characterised by an invariant measure $\mu_b$ over state space $(x,y)$ together with an entropy rate $s_{\mu_b}$ of the map. The entropy rate can be derived using the symbolic dynamics approach \cite{Eckmann85}. The following states illustrate three iterations of the map, 
\begin{equation}
\begin{split}
...y_n y_{n-1} ... y_1 y_0~.~&x_0 x_1...x_{n-1} x_n... \\
...y_{n-1} y_{n-2} ... y_0 \xi_1\overleftarrow{.} &x_1... x_n x_{n+1}... \\
...y_n y_{n-1} ... y_1 y_0\overrightarrow{.} &\xi_2 x_1...x_{n-1} x_n... \\
...y_{n+1} y_n ... y_2 y_1\overrightarrow{.} &\xi_3 \xi_2 x_1...x_{n-2} x_{n-1}... 
\end{split}
\label{eq: iterations}
\end{equation}
where $\xi_1$, $\xi_2$, and $\xi_3$, are binary symbols introduced by the stochastic variable $\xi$. Note that the probabilites for these may depend on the complete $x$ and $y$ sequences. Assume that we observe the system at a finite level of resolution with a binary precision $n+1$ as above. Then, if we have an infinite sequence of iterations, we will almost always know, by looking at the history, what is shifted in from lower levels of resolution. The only uncertainty comes from the stochastic component $\xi$ and from which direction the sequence is shifted. The latter term just contributes with a constant, $\log 2$. Assuming that the system is ergodic, which is supported by the calculations at the end of the paper for the stochastic characteristics we will use, the ergodicity theorem implies that a spatial average using $\mu_b$ can be used instead of a temporal average. Then the average rate of entropy production can be written
\begin{equation}
s_{\mu_b} = \sum \limits_{x, y} \mu_b(x,y) \sigma (q_0(x,y)) + \log 2,
\label{eq: entropy rate}
\end{equation}
where $\sigma(p)$ is the entropy function: $\sigma(p)= -p \log p - (1-p) \log(1-p)$. 

\begin{figure}[htbp]
\begin{center}
\includegraphics{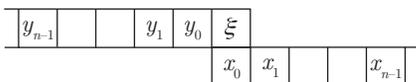}
%\epsfile{file=test-1.eps,scale=0.8}
\caption{\label{fig: configuration}{The spin block configurations needed to capture the statistics that determines the equilibrium properties of a two-dimensional spin system.}}
\end{center}
\end{figure}

Next, we present the representation of the two-dimensional spin system that has a structure identical to the one we have used for the stochastic baker's map in Eqs.~(\ref{eq: bi-infinite sequence}, \ref{eq: mirror map}). The equilibrium state in a two-dimensional spin system with nearest neighbour interactions can be characterized by probabilities of spin configurations of the form in Fig. ~\ref{fig: configuration} \cite{Kramers41, Alexandrowicz71, Katznelson72, Meirovitch77, Lindgren89, Goldstein90}. 

Denote the configuration of size $2n$ excluding the spin $\xi$ by $B_n(y_{n-1} y_{n-2} ... y_0 . x_0 x_1 ... x_{n-1})$ or shorter $B_n(x,y)$. Spin variables are 0 and 1. The conditional probability for spin $\xi$ being 0 given the configuration is denoted by $p_n(x,y)$, and it is determined by the translation invariant measure $\mu$ that characterises the equilibrium system. In order to capture the full characteristics of the spin system, one needs to consider the limit $n \rightarrow \infty$.

In this representation the relation between the stochastic baker's map and the conditional probability characterising the spin system is clear. By applying the conditional probability $p_n(x,y)$ (in the limit of $n \rightarrow \infty$) we shift out symbol $x_0$ from the $x$ sequence and we shift in symbol $\xi$ to the $y$ sequence. In this way we get a new spin configuration in the same way as we get a new position in the baker's map. This leads to the main result of this paper: If we let the probability distributions for $\xi(x,y)$ be identical to the conditional probabilities in the spin system,
\begin{equation}
q_0(x,y)=\lim _{n \to \infty}p_n(x,y),
\label{eq: baker-spin}
\end{equation}
we get a dynamics of the stochastic baker's map with an invariant measure $\mu_b$ that is equal to the translation invariant measure of the spin system, $\mu_b = \mu$.

The measure $\mu$ determines the statistical mechanics properties of the spin system. For finite block size $n$, let $\mu_n(x,y)$ denote the corresponding measure. Spin system symmetry requires $\mu_n(x,y) = \mu_n(y,x)$, and it is reflected in the symmetric construction of the stochastic baker's map in Eq.~(\ref{eq: mirror map}). The entropy $s$ (in terms of Boltzmann's constant $k_B$) of the spin system is given by 
\begin{equation}
s = \lim \limits_{n \to \infty} \sum \limits_{x, y \in \{0,1\}^n} \mu_n(x,y) \sigma (p_n(x,y)),
\label{eq: entropy}
\end{equation}
Because of the construction of $\xi$ in the stochastic baker's map, by designing it from the characteristics of the spin system, Eq.~(\ref{eq: baker-spin}), the spin system entropy is up to a constant identical to the entropy rate of the baker's map, Eq.~(\ref{eq: entropy rate}), $s=s_{\mu_b}- \log 2$. The difference between successive improvements of the entropy estimate by increasing block size $n$ can be interpreted as correlation information in the system \cite{Lindgren89}. 

Let us consider the Ising model with nearest neighbour interactions with interaction constant $J=1$ (in terms of $k_B$). Then the energy can be written 
\begin{equation}
u = 2 \mu_1(0,0)(1-2p_1(0,0))- 2 \mu_1(1,1)(1-2p_1(1,1))
\label{eq: energy}
\end{equation}
where $\mu_1(0,0)$ denotes the probability for a $B_1$ block to be (0.0), and similarly for the other term. The equilibrium state is characterized by a minimum in the free energy $g$ (in terms of $k_B$) with respect to variations in the measure $\mu$,
\begin{equation}
g = u - T s.
\label{eq: free energy}
\end{equation}

We now discuss a procedure that gives an approximation of the invariant measure $\mu_b$ of the stochastic baker's map, and consequently it also determines the translation invariant measure $\mu$ of the spin system.

If we choose $q_0(x,y)$ based on statistics from Monte Carlo simulations and run the stochastic baker's map $\phi$, does the system converge to a stationary measure that reproduces the physical properties of the spin system? One way to investigate this numerically, is to make approximations by choosing a certain resolution for positions in the unit square in the baker's map as well as a certain level of resolution for the dependence of $q_0(x,y)$ on position $(x,y)$. For the positions in the dynamics, we divide the unit square into $2^m$ equal squares, which means that we use a binary precision of $m$. This is the resolution we will use to estimate the invariant measure $\mu_b$ of the baker's map dynamics. Since the conditional probability $q_0(x,y)$ is based on the statistics over the blocks of $2n+1$ spins in Fig.~\ref{fig: configuration}, the binary precision in the spatial dependence of $q_0(x,y)$ is $n$. We require that $n \le m$.

When we consider the baker's map approximated to a certain spatial resolution $m$ it turns into a Markov process. The transitions can be represented by a finite resolution version $h_m$ of $f$, cf. Eq.~(\ref{eq: bi-infinite sequence}), 
\begin{equation}
\begin{split}
h_m& (y_{m-1} ... y_1 y_0 . x_0 x_1 ... x_{m-2} x_{m-1}) =\\
& = (y_{m-2} ... y_0 \xi . x_1 x_2 ... x_{m-1} \zeta)
\end{split}
\label{eq: baker2}
\end{equation}
Here a new stochastic variable $\zeta$ enters at the finest level of resolution in the expanding dimension since a new binary symbol is shifted in from indistinguishable positions. In order to shift in symbols in a consistent way, we use the available information that was used to form the conditional probability $q_0$. This gives us the conditional probability $z_0$ for $\zeta$ to be 0, given $(x_{m-n} ... x_{m-1})$. We include this conditional probability in $h_m$, and then we get a revised stochastic baker's map $\eta_m$ by combining $h_m$ with its mirror image, as in Eq.~(\ref{eq: mirror map}).

This means that we can calculate the invariant measure $\mu_b^{(m)}$ at resolution $m$ by considering a "master equation" for the density function $\rho_m(x, y; t)$ of the stochastic baker's map at the resolution $m$,
\begin{equation}
\rho_m(x, y; t+1)=\sum_{x',y' \in \{0,1\}^m} \rho_m(x', y'; t) P(x',y' \rightarrow x,y)
\label{eq: markov}
\end{equation}
where the transition probabilities $P$ are based on the map $\eta_m$ from Eq.~(\ref{eq: baker2}) and the discussion above. Using the double sequence representation, Eq.~(\ref{eq: markov}) can be written
\begin{equation}
\begin{split}
&\rho_m(y_n...y_1.x_1...x_n; t+1)=\\
& 
{
\begin{split} 
=\frac{1}{2}\sum_{x_0, y_{n+1}} &\rho_m(y_{n+1}...y_2.x_0...x_{n-1}; t) \times \\
&\times q_{y_1}(y_{n+1}...y_2.x_0...x_{n-1}) z_{x_n}(x_0...x_{n-1})+
\end{split} 
}\\
&
{
\begin{split} 
+\frac{1}{2}\sum_{x_{n+1}, y_0} &\rho_m(y_{n-1}...y_0.x_2...x_{n+1}; t) \times \\
&\times q_{x_1}(x_{n+1}...x_2.y_0...y_{n-1}) z_{y_n}(y_0...y_{n-1}).
\end{split}
}
\end{split}
\label{eq: markov2}
\end{equation}
Here we use the notation $q_1=1-q_0$ and $z_1=1-z_0$. 

In other words, the map $\eta_m$ defines a Markov process with a transition matrix $P$ of size $2^{2m} \times 2^{2m}$, and the invariant measure $\mu_b^{(m)}$ is then the dominating eigenvector of the $P$ matrix.

\begin{figure}[htbp]
   \centering
   \includegraphics[scale=0.7]{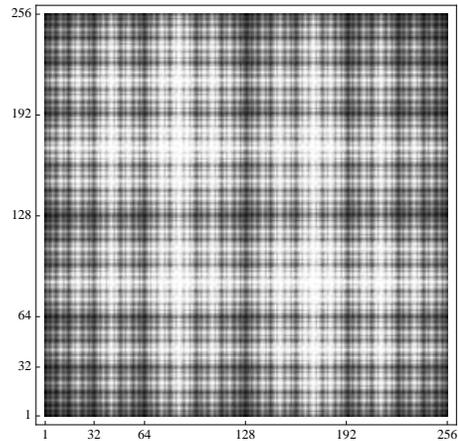}
   \caption{\label{fig: measure1}{The invariant measure $\mu_b$ of the stochastic baker's map derived at resolution $m=8$ and at temperature $T=4$, calculated as the largest eigenvector of Eq.~(\ref{eq: markov2}}). The positions $(x,y)$ in state space are numbered according to the binary expansion of the coordinates, i.e., from 0 to 255.}
\end{figure}

\begin{figure}[htbp]
\begin{center}
\includegraphics[scale=0.8]{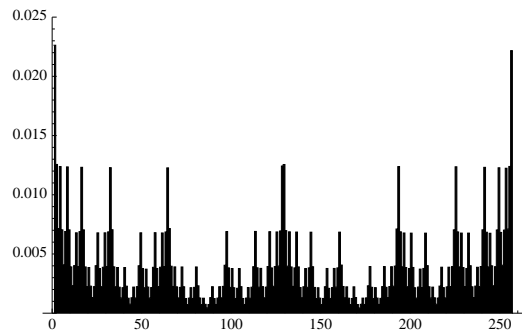}
%\epsfile{file=test-1.eps,scale=0.8}
\caption{\label{fig: projected measure1}{The invariant measure of Fig.~\ref{fig: measure1} projected onto $x$. Positions $x$ are numbered according to the binary expansion of the coordinates. 
%The highest contributions are found at 0 and 255 as they correspond to spin sequences of eight consecutive 0's and 1's respectively.
}}
\end{center}
\end{figure}

We illustrate the construction of stochastic baker's maps by using the procedure above for two temperatures in the Ising model. The conditional probabilities for $\xi$ and $\zeta$, i.e., $q$ and $z$ in Eq.~(\ref{eq: markov2}), are in the first case, for temperature $T=4$, based on statistics from Monte Carlo simulations on a $200\times 200$ lattice, in total $3.1\times 10^6$ block configurations (as in Fig.~\ref{fig: configuration}) with $n=3$. The free energy is calculated by averaging the energy and estimating the entropy as in Eqs.~(\ref{eq: entropy},\ref{eq: energy}), and the result is $g_{MC} \approx -3.03643$ (same as the exact value at this precision \cite{Onsager44}). Then we solve the invariant measure $\mu_b^{(m)}$ for the corresponding baker's map, using Eq.~(\ref{eq: markov2}) with a resolution $m=8$. The invariant measure shown in Fig.~\ref{fig: measure1}, and its projection to the $x$ axis in Fig.~\ref{fig: projected measure1}, both exhibit a fractal stucture as one may expect. We can now apply the invariant measure $\mu_b$ to Eqs.~(\ref{eq: entropy},\ref{eq: energy}) to calculate the free energy of the baker's map. The result is $g_b \approx -3.03634$, which deviates less than $10^{-4}$ from the Monte Carlo value.

In order to get a qualitative picture on how the different levels of resolution influence the construction of the map, we design a series of maps based on different values for $m$ and $n$. We choose a lower temperature, $T=2.2$ just below the critical temperature $T_c=2.269$, since correlations are stronger here and that may show up in the dependence on block size $n$. The construction of the baker's maps is now based on statistics from MC simulations, in total $3.9\times 10^7$ spin blocks with $n=5$.

\begin{figure}[htbp]
\begin{center}
\includegraphics[scale=0.65]{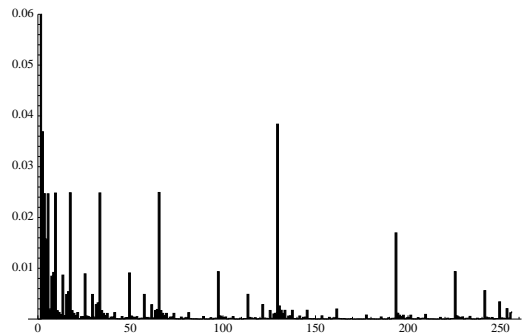}
\caption{{The invariant measure $\mu_b$ at $T=2.2$, projected onto $x$, at resolution level $m=8$. The first column, representing $x=0$ is cut off as it reaches 0.55. The temperature is below the critical level $T_c=2.269$, as is seen from the broken left-right symmetry, i.e., the symmetry between $x$ and $1-x$. (Positions are numbered as in Fig. \ref{fig: projected measure1}.)}}
\label{fig: projected measure2}
\end{center}
\end{figure}

The result is summarized in Table \ref{table}, showing that at sufficiently large blocks ($n\ge 4$) we get baker's maps that reproduce the free energy value of the MC simulation. Note that the spatial resolution of the state space (given by $m$) does not influence the free energy of the map as much as the choice of dependence on block size $n$. The invariant measure for $m=8$ and $n=4$ is shown %in Fig.~\ref{fig: measure2} 
with the projection to the $x$ axis in Fig.~\ref{fig: projected measure2}. Since we are below the critical temperature the left-right symmetry is broken. This exercise illustrates that the invariant measure of the stochastic baker's map, also at finite resolution, is sufficiently close to the measure of the spin system to reliably reproduce the free energy.

\begin{table}[htdp]
\caption{{\label{table}}{Free energy for the stochastic baker's map at different levels of resolution of state space $m$ and block size $n$ for the conditional probabilities. The rightmost column shows the Monte Carlo estimates. The exact value of the free energy with this precision is $-2.0907$ \cite{Onsager44}.}}
\begin{center}
\begin{tabular*}{0.48\textwidth}{@{\extracolsep{\fill}}c|c c c c c c c }
$n$ & $m=1$ & $m=2$ & $m=3$ & $m=4$ & $m=8$ & MC \\
\hline
1 & $-2.0807$ & $-2.0804$ & $-2.0805$ & $-2.0808$ & $-2.0815$ & $-2.1009$ \\
2 & - & $-2.0894$ & $-2.0890$ & $-2.0888$ & $-2.0887$ & $-2.0928$ \\
3 & - & - & $-2.0905$ & $-2.0904$ & $-2.0902$ & $-2.0913$ \\
4 & -& - & - & $-2.0907$ & $-2.0906$ & $-2.0909$ \\
5 & - & - & - & - & $-2.0906$ & $-2.0907$ \\
\end{tabular*}
\end{center}
\label{default}
\end{table}%

We have demonstrated that there is a class of stochastic baker's transformations that reproduce the equilibrium measure in two-dimensional spin systems with nearest neighbour interactions. The generalization to longer (but still finite interactions) is straight-forward. This means that for any two-dimensional spin system, there is a corresponding stochastic baker's map that contains the equilibrium properties of the spin system. 
The difference between our result and those obtained for coupled map lattice (CML) models, e.g., \cite{Sakaguchi99}, is that, in our approach, spin configurations (of a certain shape) are represented by positions in the unit square. This leads to the result that the process of moving across a spin lattice capturing the average physical characteristics is equivalent to a certain baker's map dynamics. In the former work, spatial configurations in the spin system are represented by corresponding spatial configurations in the CML system, and that cannot be used to establish the connection between statistical mechanics and dynamical systems that we find in the present study. 

An interesting question is whether the finite resolution map of Eqs.~(\ref{eq: baker2},\ref{eq: markov}) can be used in a variational approach to find the invariant baker's map measure that minimizes the free energy. Such a procedure would have strong connections to variational methods \cite{Kikuchi51, Goldstein90, Schlijper90}. The reformulation of the variational problem for the spin system into a dynamical systems problem may offer new perspectives based on dynamical systems theory. Work along these lines is in progress. 

The author thanks Martin Nilsson Jacobi for several constructive discussions, and both him and Kolbj\o rn Tunstr\o m for valuable comments on the manuscript.

\bibliography{articles} 

\end{document}